\documentclass{elsart}
\usepackage{amsmath}
\usepackage{rotating}

\input{tcilatex}

\begin{document}

\begin{frontmatter}

\title{On the integrated behaviour of non-stationary volatility in stock markets}

\author{Andreia Dionisio*, Rui Menezes** and Diana A. Mendes**}

\address{*University of Evora, Center of Business Studies, CEFAGE-UE, Largo Colegiais, 2, 7000 Evora, Portugal, E-mail: andreia@uevora.pt;
 **ISCTE, Av. Forcas Armadas, 1649-025 Lisboa, Portugal }

\begin{abstract}
This paper analyses the behaviour of volatility for several international stock market indexes, namely the SP 500 (USA), the Nikkei (Japan), the PSI 20 (Portugal), the CAC 40 (France), the DAX 30 (Germany), the FTSE 100 (UK), the IBEX 35 (Spain) and the MIB 30 (Italy), in the context of non-stationarity. Our empirical results point to the evidence of the existence of integrated behaviour among several of those stock market indexes of different dimensions. It seems, therefore, that the behaviour of these markets tends to some uniformity, which can be interpreted as the existence of a similar behaviour facing to shocks that may affect the worldwide economy. Whether this is a cause or a consequence of market globalization is an issue that may be stressed in future work.
\end{abstract}


\end{frontmatter}

\section*{Introduction}

The persistence of stock price volatility is a well-known stylized fact in
the financial literature. Much of the empirical tests of volatility
presented in the literature rely on the standard GARCH approach proposed by
Bollerslev and Wooldrigde (1992), and often produce evidence that the
conditional volatility is highly persistent. The stock prices volatility
also presents some attributes that are typically non-stationary, an issue
that requires the consideration of a special class of conditional
heteroskedasticity models based on the IGARCH specification proposed by
Engle and Bollerslev (1986). Under this specification, there is no need to
differentiate the series when they prove to be non-stationary in order to
apply the conditional heteroskedasticity models, thus retaining the richness
of information contained in the original series.

The main purpose of this paper is to compare the volatility between several
international stock market indexes, namely the S\&P 500 (USA), the Nikkei
(Japan), the\ Hang-Seng (Hong-Kong), the PSI 20 (Portugal), the CAC 40
(France), the DAX 30 (Germany), the FTSE 100 (UK), the IBEX 35 (Spain), the
ASE (Greece) and the MIB 30 (Italy), in the context of non-stationarity. We
use the daily closing prices of these indexes to perform the tests and to
present the empirical results.

In this paper we applied Johansen tests (Johansen, 1988) in order to test
cointegration between non-stationary variables, along with tests for weak
exogeneity (Johansen and Juselius, 1990). The results were then compared to
those obtained by the Granger causality tests in order to get evidence on
strong exogeneity of the variables. Besides, stochastic integrated
conditional heteroskedasticity specifications based on IGARCH and FIGARCH
(Meddahi and Renault, 2004) models were also attempted in order to capture
the likely non-stationary attribute of the series under the context of
conditional volatility.

Our empirical results show evidence of existence of integrated behaviour
among several stock market indexes of different dimensions. It seems,
therefore, that the behaviour of these markets tends to some uniformity,
which can be interpreted as the existence of a similar behaviour facing to
shocks that may affect the worldwide economy.

The rest of the paper is organized as follows. In Section 1 we present a
brief discussion of the background theory. Section 2 discusses the results
of testing for the long-run relationship in stock indexes using
cointegration techniques. Next, we present in Section 3 the results of
fractional volatility in stock returns using GARCH, IGARCH and FIGARCH
specifications. Finally, Section 4 presents the conclusions.

\section{Background Theory}

The standard GARCH framework (Bollerslev \emph{et al.}, 1992) often produces
evidence that the conditional volatility process is highly persistent and
possibly not covariance-stationary, suggesting that a model in which shocks
have a permanent effect on volatility might be more appropriate. This is a
property of the integrated GARCH (IGARCH) model (Engle and Bollerslev, 1986)
which has infinite memory.

Following Engle (1982), we consider the time series $y_{t}$ with the
associated error 
\begin{equation}
e_{t}=y_{t}-E_{t-1}y_{t}
\end{equation}%
where $E_{t-1}$ is the expectation operator conditioned on time $t-1$. A
generalized autoregressive conditional heteroskedasticity (GARCH) model
where 
\begin{equation}
e_{t}=z_{t}\sigma _{t},\text{ \ \ }z_{t}\sim N\left( 0,1\right)
\end{equation}%
is defined as 
\begin{equation}
\sigma _{t}^{2}=w+\alpha \left( L\right) e_{t}^{2}+\beta \left( L\right)
\sigma _{t}^{2},  \label{garch}
\end{equation}%
where $w>0$, and $\alpha \left( L\right) $ and $\beta \left( L\right) $ are
polynomials in the lag operator $L\left( L^{i}x_{i}=x_{t-i}\right) $ of
order $q$ and $p$ respectively. Expression $\left( \ref{garch}\right) $ can
be rewritten as the infinite-order ARCH process,%
\begin{equation}
\Phi \left( L\right) e_{t}^{2}=w+\left[ 1-\beta \left( L\right) \right]
v_{t},
\end{equation}%
where $v_{t}\equiv e_{t}^{2}-\sigma _{t}^{2}$ and $\Phi \left( L\right) =%
\left[ 1-\alpha \left( L\right) -\beta \left( L\right) \right] .$ One
limitation of this process applied to financial data, is that GARCH model
has short-memory model because volatility shocks decay at a fast geometric
rate. So, a way to represent the observed persistence of volatility on the
rate of returns is to approximate a unit root, resulting from that the
integrated GARCH (IGARCH) model. The specification of the IGARCH model is%
\begin{equation}
\Phi \left( L\right) \left( 1-L\right) e_{t}^{2}=w+\left[ 1-\beta \left(
L\right) \right] v_{t}.  \label{igarch}
\end{equation}

Perron (1989) demonstrated that standard tests tend to under-reject the null
of unit-root in the presence of structural breaks in the mean. In a similar
way, breaks in the conditional variance could lead to spuriously high
estimates of its degree of persistence. However, and according to Vilasuso
(2002) the IGARCH model is not an entirely satisfactory description of the
rate of returns volatility because one property of the model is infinite
memory.

Motivated by the presence of apparent long-memory in the autocorrelations of
squared or absolute returns of various financial assets, Baillie \emph{et al.%
} (1996) have introduced the fractionally integrated GARCH, the FIGARCH
model. Analogously for the ARFIMA ($k,d,l$) process for the mean described
by 
\begin{equation}
a\left( L\right) \left( 1-L\right) ^{d}y_{t}=b\left( L\right) e_{t},
\end{equation}%
where the $a\left( L\right) $ and $b\left( L\right) $ are polynomials in the
lag operator of order $k$ and $l$ and $e_{t}\sim N\left( 0,1\right) $. The
FIGARCH ($p,d,q$) process for $\left\{ e_{t}\right\} $ is defined by%
\begin{equation}
\Phi \left( L\right) \left( 1-L\right) ^{d}e_{t}^{2}=w+\left[ 1-\beta \left(
L\right) \right] v_{t}  \label{figarch}
\end{equation}%
where $0\leq d\leq 1$ is the fractional difference parameter.

The primary purpose of this approach is to develop a more flexible class of
processes for the conditional variance that are more capable of explaining
and representing the observed temporal dependencies in financial market
volatility. For example, the special cases: $d=0$ corresponds to modelling \
a GARCH process and $d=1$ corresponds to an IGARCH process. For d\TEXTsymbol{%
>}0 the process is long-memory. The ARFIMA model essentially disentangles
the short-run and the long-run dynamics by modelling the short-run behaviour
through the conventional ARMA lag polynomials $\left[ a\left( L\right) \text{
and }b\left( L\right) \right] $ while the long-run characteristic is
captured by the fractional differencing parameter $\left( d\right) .$

The FIGARCH process combines many of the features of the fractionally
integrated process for the mean [ARFIMA$(k,d,l)$ process] together with the
regular GARCH process for the conditional variance. It implies a slow
hyperbolic rate of decay for the lagged squared innovations in the
conditional variance function, although the cumulative impulse response
weights associated with the influence of volatility shocks on the optimal
forecasts of the conditional variance tend to zero (Baillie \emph{et al}.,
1996).

The common approach for estimation of ARCH models, assumes a conditional
normality of the process. Under this assumption, Maximum Likelihood
Estimates (MLE) for the parameters of FIGARCH($p,d,q$) can be considered the
most efficient estimation process. For the FIGARCH($p,d,q$) model with $d>0$
the population variance is not finite. However, subject to the regularity
conditions specified by Baillie \emph{et al.} (1996), conditioned on the
pre-sample values will not affect the asymptotic distributions of the
resulting estimators and test statistics. In most practical applications
using financial data, the standardized innovations $z_{t}=e_{t}\sigma
_{t}^{-1}$ are leptocurtik and not \emph{i.i.d}. normally distributed
through time. In this situations, Baillie \emph{et al.} (1996) point to the
use of the robust \emph{Quasi-MLE} (QMLE).

\section{Testing for long-run relationships in stock indexes}

We start our empirical analysis by testing for long-run equilibrium
relationships between the stock indexes variables used in our study. The
analysis is based on cointegration techniques using autoregressive (VAR)
systems in order to ascertain the extent of integration in these stock
market indexes. Our goal is to identify common behaviour and dependencies in
these markets, regardless of the occurrence of peaks, slumps, or periods of
price stability.

The integration of international capital markets has received a large amount
of attention from financial researchers over the past few years, mainly
because of factors such as the relaxation of exchange controls and increased
international information flows, technological developments in
communications and trading systems, and the introduction of innovative
financial products.

Cointegration tests provide useful information for strategic asset
allocations. One of the arguments in favour of the international
diversification is that it lowers portfolio risks without sacrificing
expected returns on the presumption that world stock prices are independent.
Examples of some recent studies on this topic include Darrat, Elkhal and
Halkim (2000) and Phylaktis and Ravazzolo (2002).

In fact, the literature on stock markets has been mainly focused on
international portfolio diversification.

An increased correlation between stock market indexes is usually interpreted
as a rise in the extent of market integration. This leads to a higher
tendency for a shock in one country being transmitted to another one. But a
straightforward use of this approach may sometimes give misleading
conclusions, because of the non-stationary nature of most stock market price
variables.

The recognition of the importance of the nonstationarity property of stock
prices led some researchers to explore possible long-run relations among
national and international stock markets using the notion of cointegration
as defined by Engle and Granger (1987).

The data set used in this paper contains 4221 daily closing prices spanning
the period from January, 5, 1990 to April, 7, 2006 for 10 international
stock indexes, namely he S\&P 500 (USA), the Nikkei (Japan), the\ Hang-Seng
(Hong-Kong), the PSI 20 (Portugal), the CAC 40 (France), the DAX 30
(Germany), the FTSE 100 (UK), the IBEX 35 (Spain), the ASE (Greece) and the
MIB 30 (Italy).\footnote{%
All the variables are transformed into natural logarithms.} Figure 1
presents the general behaviour and trend of the stock indexes covering the
period under study.

\FRAME{ftbphFU}{5.4267in}{3.621in}{0pt}{\Qcb{Behaviour and trend of the
natural logarithm of the stock indexes. January, 5, 1990 \ - April, 7, 2006.}%
}{\Qlb{stock prices}}{dionisiofig1.eps}{\special{language "Scientific
Word";type "GRAPHIC";maintain-aspect-ratio TRUE;display "USEDEF";valid_file
"F";width 5.4267in;height 3.621in;depth 0pt;original-width
6.0027in;original-height 3.9946in;cropleft "0";croptop "1";cropright
"1";cropbottom "0";filename '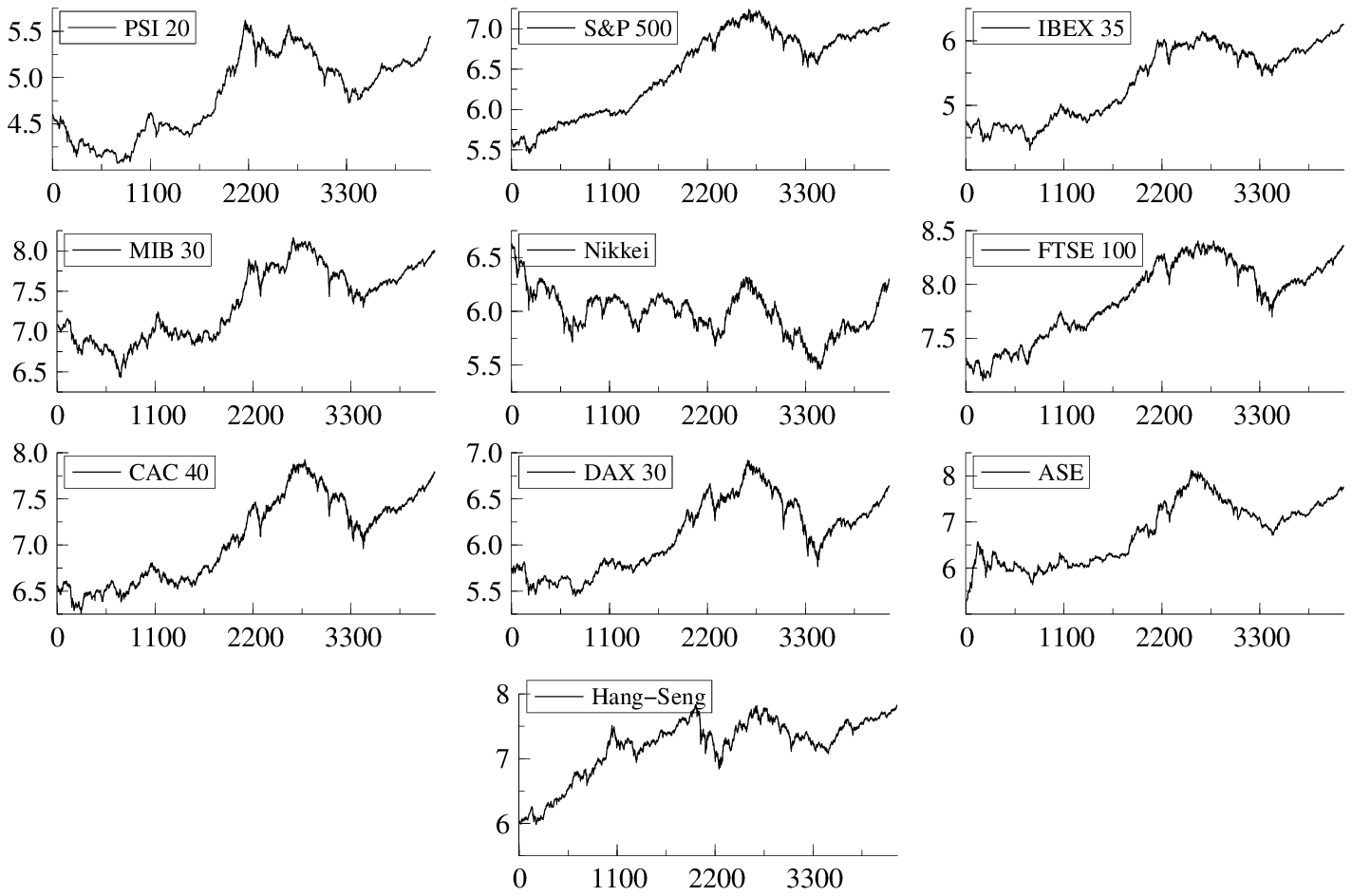';file-properties "XNPEU";}}

As may be seen, there are similarities in the behaviour and general trend in
many of the series presented in Figure \ref{stock prices}, namely the PSI
20, IBEX 35, CAC 40, FTSE 100, DAX 30, MIB 30 and S\&P 500, although they
may differ in the short-run. The statistical results reveal that all stock
indexes are strongly leptocurtik, non-normally distributed and exhibit
evidence of hereroskedasticity.

The ADF and\ the Elliott-Rothenberg-Stock tests for unit-roots were applied.
The unit-root hypothesis was not rejected at standard significance levels in
any case, for the series in levels, independently of the inclusion of a
constant term and a deterministic trend in the ADF and ERS regressions. On
the other hand, for first differences, the null hypothesis of a unit root
was strongly rejected, indicating that each of the first-differenced series
is stationary.

In order to evaluate the possible cointegration in these stock indexes, we
applied\ Johansen tests (Johansen, 1988) for cointegration between these
non-stationary variables. Firstly, we tested for bivariate cointegration in
each pair of stock indexes. We performed this test for 45 different pairs of
stock indexes. We also computed weak exogeneity tests and Granger causality
tests.

The finding of a cointegrating vector between pairs of series indicates that
over the sample the series move together in an equilibrium relationship. The
term equilibrium in the cointegration literature is sometimes synonymous
that the series maintain a constant relationship throughout the sample. It
does not mean that over specific sub-periods the series did not move apart.

The main results of these tests point to the existence of 18 pairs of stock
indexes that show signs of cointegration. Table 1 presents the results of
the cointegration tests and the exogeneity tests for the pairs of stocks
that show evidence of statistical significant bivariate cointegration.%
\footnote{%
Under the null hypothesis of cointegration, $r=0$ corresponds to the case
where there are no cointegrating vectors, and $r<=1$ corresponds to the case
where there is at most one cointegrating vector [column (1)]. The next three
columns provide information about the eigenvalues and the trace and maximum
eigenvalue test statistics for each hypothesis. Columns (5) and (6) give the
results of the exogeneity tests, where the parameter denoting the speed of
adjustment to long-run equilibrium is tested for being zero in the null
hypothesis.}

\bigskip

\begin{center}
[Please insert DionisioFigure2 here]

\bigskip
\end{center}

The results of the weak exogeneity tests are in conformity with the results
obtained with the Granger causality tests,\footnote{%
The Granger causality results are available upon request to the authors.}
leading us to conclude that there exists strong exogeneity in the reported
cases.

Our results also show that there is a close integration within the European
stock markets and also between some European stocks and the US and the Asian
stock markets. Several factors may explain this situation. One of these
factors is that the companies around the world strongly exposed to the
global business cycle, leading the national stock markets to move together
more tightly [Barton \emph{et al}. (2002)].\footnote{%
We also performed multivariate cointegration tests for all the stock indexes
contained in our database. We found for the entire set of our variables 3
cointegrating vectors. The speed of adjustment to the long-run equilibrium
relationship is statistically significant for the PSI 20 (0.002), MIB 30
(0.001), DAX 30 (0.002), ASE (0.003) and CAC 40\ (0.001) indexes. It seems
therefore that there is a stronger interaction among the European
continental indexes (except the IBEX 35), which goes in same direction of
the macroeconomic behaviour of the underlying economies.
\par
The weak exogeneity tests for the whole set of indexes showed no evidence of
exogeneity in\ the PSI 20, DAX 30 and ASE indexes. This may indicate that
the corresponding stock markets can receive more influences from the other
stock markets, being more exposed to shocks in the global economy than other
stock markets.}

\section{Volatility in stock returns}

\bigskip

We now turn to consider the fractional property of the stock returns
volatility. We fit the conditional hereroskedasticity models using the
estimation method proposed by Chung (2001) based on \emph{Quasi--Maximum
Likelihood Estimation} (QMLE) methods: GARCH ($1,1$), IGARCH ($1,1$) and
FIGARCH ($1,d,1$).The rate of returns of the stock indexes was computed as
follows:%
\begin{equation}
R_{i,t}=\ln \left( \frac{P_{i,t}}{P_{i,t-1}}\right) =\mu +R_{i,t-1}+e_{i,t},
\end{equation}%
where $P_{i,t}$ is the value of the underlying stock index $i$ at time $t,$ $%
\mu $ is a constant and $e_{i,t}\sigma _{i,t}^{-1}$ is $i.i.d.\sim N\left(
0,1\right) ,$ using the model%
\begin{equation}
\sigma _{i,t}^{2}=w+\beta _{1}\sigma _{i,t-1}^{2}+\left[ 1-\beta _{1}\left(
L\right) -\left( 1-\Phi \right) \left( 1-L\right) ^{d}\right] e_{i,t}^{2}.
\end{equation}

The results displayed in Table 2 show that for all stock index returns the
parameter $d$ of the FIGARCH model is always statistically significant, with
values between 0.3 and 0.5. On the other hand, the GARCH model, which
assumes that $d=0,$ produces a lower log-likelihood statistic most of the
times. We can also see that the estimated GARCH (1,1) parameters do not
differ much from the estimated IGARCH (1,1). The results show that the
dynamics of the conditional variance of stock index returns are best
represented by the FIGARCH model.

Our results indicate that the residuals $\left( e_{i,t}\right) $ are not
stationary, since for all cases we reject the hypothesis that $d=0$ while at
the same time we can see that it is precisely the FIGARCH model, where $d<1$
that presents the highest values for the log-likelihood parameter. This
seems to reveal that the IGARCH model is not the best alternative to
estimate the volatility of the index rate of returns. This result seems to
indicate that the stock index series are not $I$(1) and the first
differences are not $I$(0), and so the main conclusions and results obtained
with the traditional stationary tests and cointegration analysis could be
misleading.

\bigskip

\begin{center}
[Please insert DionisioFigure3 here]

\bigskip
\end{center}

{\scriptsize Note: The conditional mean of each rate of return is modelled
as a constant }$\mu ,${\scriptsize \ and }$w${\scriptsize \ is constant in
the conditional variance. The values in brackets refer to the
standard-deviation. For the IGARCH (1,1), the estimation of }$\Phi $%
{\scriptsize \ is unbounded.}

It is important to note that the European stock indexes, whose underlying
economies are more developed (CAC 40, DAX 30 and FTSE 100) present the
highest levels for the parameter $d$, with values of, respectively, 0.5158,
0.5494 and 0.5135. Since for these index returns $d>0.5$, there are signals
of mean-reverting behaviour, which seems to contradict the efficient market
hypothesis.

On the other hand, the stock markets characterized by lower dimension and
levels of liquidity, such as PSI 20, ASE, MIB 30 and Hang-Seng as well as
the S\&P 500 and the Nikkei (which are well developed markets), present
values of $d$ smaller than 0.5, which means that the corresponding return
series are stationary. Given these results, we cannot conclude that the
dimension of the market and its level of liquidity are factors that can
promote stationarity on the volatility, being closer to the efficient market
hypothesis.

\section{Conclusions}

In this paper we study the dynamics of the stock price indexes and of the
rate of returns volatility.

For the cointegration long-run relationships the results obtained are mixed,
with some market indexes being bivariately cointegrated and pothers not. Out
of the 45 bivariate models tested, 18 show signs of cointegration. This is
the case of most European markets, and also the case of the S\&P 500 with
some of the other markets. In the Europe, this is especially so among some
continental stock indexes, namely the PSI 20, DAX 30, ASE, MIB 30 and FTSE
100, in which the ASE and MIB 30 are essentially endogenous variables and
the DAX 30, FTSE 100 and PSI 20 present weak exogeneity.

Relating to the volatility analysis, the results show that the FIGARCH model
is better suited to capture the behaviour of stock indexes returns than the
commonly used GARCH model and also the IGARCH model. All stock index returns
exhibit evidence of fractional integration.

Indeed, we found signs of long-memory effects, which seems to indicate that
there is evidence of dynamics, not only in the dimension of prices, but also
on the volatility. The finding of a significant integration parameter $0<d<1$
seems to indicate that the original series of stock index prices are not
integrated of order 1 $\left[ I(1)\right] $, being around $I$(1.5). This
result indicates that the conventional "linear" cointegration analysis may
pose some problems since it is based on the assumption that the time series
are $I$(1) and the residuals resulting from the estimation of a long-run
equilibrium model are stationary $\left[ I(0)\right] $.

A possible alternative would be a model of fractional cointegration, which
is however out of the scope of this research work.

Our empirical results point to the evidence of the existence of integrated
behaviour among several stock market indexes of different dimensions. It
seems, therefore, that the behaviour of these markets tends to some
uniformity, which can be interpreted as the existence of a similar behaviour
facing to shocks that may affect the worldwide economy.

\end{document}